\documentstyle[llncs]{article}

\def\ll{{\langle}}
\def\rl{{\rangle}}
\def\bt{{\bf t}}
\def\rules{{\Delta}}
\def\grammar{{\sf G}}
\def\hd{{\it head}}
\def\hds{{\it heads}}
\def\tl{{\it tail}}
\def\tls{{\it tails}}
\def\exp{{\it expand}}
\def\exps{{\it expands}}
\def\sel{{\it selects}}
\def\open{{\it open}}
\def\opens{{\it opens}}
\def\incl{{\it subset}}
\def\inclv{{\it subsetv}}

\newcommand\sem[1]{{[\!\![#1]\!\!]}_{\grammar}}
\newcommand\tsem[1]{{\ll\!\!\ll#1\rl\!\!\rl}_{\grammar}}
\newcommand {\definedas}{\stackrel{\em def}{=}} 
\newcommand {\first}[1]{{\it #1}} 
\def\terms{{\cal T} }

\def\true{{\sl true} } 
\def\false{{\sl false}} 
\def\lor{\vee} 
\def\land{\wedge}

\begin{document} 

\Large

\bibliographystyle{plain}

\title{On Dart-Zobel Algorithm for Testing Regular Type Inclusion} 
\author{Lunjin Lu and John G. Cleary}
\institute{ Department of Computer Science\\ University of Waikato \\
Hamilton, New Zealand\\ Phone: +64-838-4627/4378\\
\{lunjin,jcleary\}@cs.waikato.ac.nz}

\date{}

\maketitle
 
\begin{abstract}  
This paper answers open questions about the correctness and the
completeness of Dart-Zobel algorithm for testing the inclusion
relation between two regular types. We show that the algorithm is
incorrect for regular types. We also prove that the algorithm is
complete for regular types as well as correct for tuple distributive
regular types. Also presented is a simplified version of Dart-Zobel
algorithm for tuple distributive regular types.

\noindent
{\bf Keywords:} type, regular term language, regular term grammar, 
                tuple distributivity
\end{abstract}

\section{Introduction}
Types are ubiquitous in programming
languages~\cite{Cardelli:Wegner:85}. They make programs easier to
understand and help detect errors since a large number of errors
are type errors.  Types have been introduced into logic
programming in the forms of type checking and
inference~\cite{Beierle:ICLP95,Dart:Zobel:JLP92,FruhwirthSVY:LICS91,Mycroft:OKeefe:84,Yardeni:Shapiro:91}
or type
analysis~\cite{Mishra:84,Zobel:87,HeintzeJ90,HeintzeJ92,GallagherW94,Lu95,CL96,LuJLP98}
or typed
languages~\cite{Hanus:TCS,Jacods:PLDI90,Reddy:NACLP90,YardeniFS:ICLP91}.
Recent logic programming systems allow the programmer declare
types for predicates and type errors are then detected either at
compile time or at run time. Even in early logic programming
systems, built-in predicates are usually typed and type checking
for these predicates are performed at run time.  The reader is
referred to \cite{Pfenning92} for more details on type in logic
programming.

A \first{type} is a possibly infinite set of ground terms with a
finite representation. An integral part of any type system is its
type language that specifies which sets of ground terms are
types. To be useful, types should be closed under intersection,
union and complement operations. The decision problems such as the
emptiness of a type, inclusion of a type in another, equivalence
of two types should be decidable.  Regular term
languages~\cite{GecsegS84,ComonDGLTT98}, called regular types,
satisfy these constraints and has been used widely used as
types~\cite{Soloman78,Mishra:84,Zobel:87,Dart:Zobel:JLP92,HeintzeJ90,Jacods:PLDI90,Reddy:NACLP90,YardeniFS:ICLP91,FruhwirthSVY:LICS91,Yardeni:Shapiro:91,HeintzeJ92,GallagherW94,Lu95,CL96,LuJLP98}.

Most type systems use \first{tuple distributive regular types}
which are strictly less powerful than regular
types~\cite{Soloman78,Mishra:84,Zobel:87,HeintzeJ90,Jacods:PLDI90,Reddy:NACLP90,YardeniFS:ICLP91,FruhwirthSVY:LICS91,Yardeni:Shapiro:91,HeintzeJ92,GallagherW94,Lu95,CL96,LuJLP98}. Tuple
distributive regular types are regular types closed under tuple
distributive closure. Intuitively, the tuple distributive closure
of a set of terms is the set of all terms constructed recursively
by permuting each argument position among all terms that have the
same function symbol~\cite{Yardeni:Shapiro:91}.  Tuple
distributive regular types are discussed in
section~\ref{sec:tuple}.

To our knowledge, Dart and Zobel's work~\cite{DartZ92} is the only
one to present, among others, an inclusion algorithm for regular
types with respect to a given set of type definitions without the
tuple distributive restriction. Set-based analysis can also be
used to deriving types based on set constraint
solving~\cite{AikenW92,AikenL94,heintze-set,heintze-decision,DevienneTT98}.
However, set constraint solving methods are intended to infer
descriptive types~\cite{Reddy:NACLP90} rather than for testing
inclusion of a prescriptive type~\cite{Reddy:NACLP90} in
another. Therefore, they are useful in different settings from
Dart-Zobel algorithm.  Dart-Zobel algorithm has been used in type
or type related analyses~\cite{Dart:Zobel:JLP92,DebrayLH97}.
However, the completeness and the correctness of the algorithm are
left open. This paper provides answers to these open questions. We
show that the algorithm is incorrect for regular types. We also
prove that the algorithm is complete for regular types in general
as well as correct for tuple distributive regular types. These
results lead to a simplified version of Dart-Zobel algorithm that
is complete and correct for tuple distributive regular types.

The remainder of this paper is organised as
follows. Section~\ref{sec:regular} defines regular types by regular
term grammars. Section~\ref{sec:algo} recalls Dart-Zobel algorithm
for testing if a regular type is a subset of another regular
type. Section~\ref{sec:open} addresses the completeness and the
correctness of their algorithm that have been left open. In
section~\ref{sec:tuple}, we show that their algorithm is both complete
and correct for tuple distributive regular types and provides a
simplified version of their algorithm for tuple distributive regular
types.

\section{Regular types} \label{sec:regular}

Several equivalent formalisms such as tree
automata~\cite{GecsegS84,ComonDGLTT98}, regular term
grammars~\cite{GecsegS84,ComonDGLTT98}, regular unary logic
programs~\cite{Yardeni:Shapiro:91} have been used to describe regular
types. In~\cite{DartZ92}, Dart and Zobel use regular term grammars to
describe regular types that are sets of ground terms over a ranked
alphabet $\Sigma$.

A regular term grammar is a tuple $\grammar=\ll
\Pi,\Sigma,\rules\rl$ where\footnote{A start symbol is not needed
in our setting.}
\begin{itemize}
\item $\Sigma$ is a fixed ranked alphabet. Each symbol in $\Sigma$
      is called a function symbol and has a fixed arity. It is
      assumed that $\Sigma$ contains at least one constant that is
      a function symbol of arity $0$.
\item $\Pi$ is a set of symbols called nonterminals. These terminals
      will be called type symbols as they represent types. Type
      symbols are of arity $0$. It is assumed that
      $\Pi\cap\Sigma=\emptyset$.
\item $\rules$ is a set of production rules of the form
      $\alpha\rightarrow\tau$ with $\alpha\in\Pi$ and
      $\tau\in\terms(\Sigma\cup\Pi)$ where $\terms(\Sigma\cup\Pi)$ is
      the set of all terms over $\Sigma\cup\Pi$. Terms in
      $\terms(\Sigma\cup\Pi)$ will be called pure type terms.
\end{itemize} 

\begin{example} \label{ex1} Let $\Sigma=\{0,s(),nil,cons(,)\}$ and 
                    $\Pi=\{Nat,NatList\}$.
                $\grammar=\ll\Pi,\Sigma,\rules\rl$ defines
                natural numbers and lists of natural numbers where
\[ \Delta= \left\{\begin{array}{l}
                   Nat \rightarrow 0,\\
                   Nat \rightarrow s(Nat),\\
                   Natlist \rightarrow nil,\\
                   Natlist \rightarrow cons(Nat,Natlist)
                  \end{array}
           \right\}
\]

{\noindent $\Box$}
\end{example}

The above presentation is slightly different from \cite{DartZ92} where
production rules with the same type symbol on their lefthand sides are
grouped together and called a type rule. For instance, production
rules in the above examples are grouped into two type rules
$Nat\rightarrow \{Nat,Natlist\}$ and
$Natlist\rightarrow\{nil,cons(Nat,Natlist)\}$.

Types denoted by a pure type term is given by a rewrite rule
$\Rightarrow_{\grammar}$ associated with
$\grammar$. $t\Rightarrow_{\grammar}s$ if $\rules$ contains a rule
$\alpha\rightarrow\tau$, $\alpha$ occurs in $t$ and $s$ results from
replacing an occurrence of $\alpha$ in $t$ by $\tau$.  Let
$\Rightarrow_{\grammar}^{*}$ be the reflexive and transitive closure
of $\Rightarrow_{\grammar}$.  The type denoted by a pure type term
$\tau$ is defined as follows.
\[ \sem{\tau} \definedas
   \{t\in\terms(\Sigma)~|~\tau\Rightarrow_{\grammar}^{*} t\}
\]
$\sem{\tau}$ is the set of terms over $\Sigma$ that can be derived
from $\tau$ by repeatedly replacing the lefthand side of a rule in
$\rules$ with its righthand side.

\begin{example} Let $\grammar$ be the regular term 
                grammar in example~\ref{ex1}. We have 
\begin{eqnarray*}
 Natlist & \Rightarrow_{\grammar} & cons(Nat,Natlist) \\
         & \Rightarrow_{\grammar} & cons(s(Nat),Natlist)\\
         & \Rightarrow_{\grammar} & cons(s(0), Natlist)\\
         & \Rightarrow_{\grammar} & cons(s(0), nil)
\end{eqnarray*} 
Thus, $\sem{Natlist}$ contains $cons(s(0), nil)$.
\end{example}  

The type represented by a sequence $\psi$ of pure type terms and a set
$\Psi$ of sequences of pure type terms are defined as follows. 
\begin{eqnarray*} 
\sem{\epsilon} & \definedas &\{\epsilon\}\\
\sem{\ll\tau\rl+\psi'} & \definedas & \sem{\tau}\times\sem{\psi'}\\ 
\sem{\Psi} & \definedas & \bigcup_{\psi\in\Psi}\sem{\psi}
\end{eqnarray*}
where $\epsilon$ is the empty sequence, $+$ is the infix sequence
concatenation operator, $\ll\tau\rl$ is the sequence consisting of
the pure type term $\tau$ and $\times$ is the Cartesian product
operator.

The set $\Pi$ of nonterminals in Dart and Zobel's type language also
contains constant type symbols. Constant type symbols are not defined
by production rules and they denote constant types. In particular,
$\Pi$ contains $\mu$ denoting the set of all terms over $\Sigma$ and
$\phi$ denoting the empty set of terms.  We will leave out constant
type symbols in this paper in order to simplify
presentation. Re-introducing constant type symbols will not affect the
results of the paper.

Dart-Zobel algorithm works with simplified regular term grammars. A
regular term grammar $\grammar=\ll\Pi,\Sigma,\rules\rl$ is simplified
if $\sem{\alpha}\neq\emptyset$ for each $\alpha\in\Pi$ and
$\tau\not\in\Pi$ for each $(\alpha\rightarrow\tau)\in\rules$. Every
regular grammar can be simplified. 

\section{Dart-Zobel Inclusion Algorithm} \label{sec:algo}
This section recalls Dart and Zobel's inclusion algorithm for
regular types. As indicated in section~\ref{sec:regular}, we shall
disregard constant type symbols and simplify their algorithm
accordingly. We note that without constant type symbols, many
functions in their algorithm can be greatly simplified.  In place
of a type rule, we use the corresponding set of production
rules. These superficial changes don't change the essence of the
algorithm but facilitate the presentation.  We shall assume that
$\grammar$ is a simplified regular term grammar and omit
references to $\grammar$ where there is no confusion.

We first describe the ancillary functions used in their algorithm. Let
$\psi=\tau_1\tau_2\cdots\tau_n$ be a non-empty sequence of pure type
terms and $\Psi$ be a set of non-empty sequences of pure type terms.
$\hd(\psi)\definedas\tau_1$ and
$\tl(\psi)\definedas\tau_2\cdots\tau_n$. $\hds$ and $\tls$ are defined
as $\hds(\Psi)\definedas\{\hd(\psi)~|~\psi\in\Psi\}$ and
$\tls(\Psi)\definedas\{\tl(\psi)~|~\psi\in\Psi\}$. The function $\exp$
rewrites a non-empty sequence into a set of sequences when necessary.
\[ \exp(\psi) \definedas 
    \left\{\begin{array}{ll} \{\psi\} & \mbox{if
     $\hd(\psi)\not\in\Pi$}\\
     \{\ll\tau\rl+\tl(\psi)~|~(\hd(\psi)\rightarrow\tau)\in\rules\} &
     \mbox{if $\hd(\psi)\in\Pi$}\\ \end{array}\right.
\]
$\exps(\Psi)\definedas\bigcup_{\psi\in\Psi}\exp(\psi)$.

The function $\sel(\tau,\Psi)$ defined below applies when $\tau$ is
pure type term and $\tau\not\in\Pi$ and $\Psi$ is a set of non-empty
sequences with $\hds(\Psi)\cap\Pi=\emptyset$. The output of
$\sel(\tau,\Psi)$ is the set of the sequences in $\Psi$ that have the
same principal function symbol as $\tau$.
\[\sel(f(\tau_1,\cdots,\tau_n),\Psi)\definedas 
  \{\psi\in\Psi~|~\hd(\psi)=f(\omega_1,\cdots,\omega_n)\}
\] Note that $f(\tau_1,\cdots,\tau_n)$ is a constant when $n=0$.

The function $\open(\psi')$ defined below applies when $\psi'$ is a
non-empty sequence with $\hd(\psi')\not\in\Pi$. $\open(\psi')$
replaces the head of $\psi'$ with its arguments.
\[ open(f(\tau_1,\cdots,\tau_n)+\psi) \definedas 
   \tau_1\tau_2\cdots\tau_n+\psi
\] 
When $n=0$, $\open(f(\tau_1,\cdots,\tau_n)+\psi)=\psi$.  Without
constant type symbols, $open$ doesn't need an extra argument as
in~\cite{DartZ92} that is used to test membership of a term in a
constant type and to indicate the required number of arguments when
the constant type symbol is $\mu$. $\opens(\Psi)\definedas
\{\open(\psi)~|~\psi\in\Psi\}$.

The inclusion algorithm $\incl(\tau_1,\tau_2)$ takes two pure type
terms $\tau_1$ and $\tau_2$ and is intended to decide if
$\sem{\tau_1}\subseteq\sem{\tau_2}$ is true or false.  The core
part $\inclv$ of the inclusion algorithm takes a sequence $\psi$
of pure type terms and a set $\Psi$ of sequences of pure type
terms that are of the same length as $\psi$ and is intended to
decide if $\sem{\psi}\subseteq\sem{\Psi}$. $\inclv$ takes a third
argument $C$ to ensure termination. $C$ is a set of pairs
$\ll\beta,\Upsilon\rl$ where $\beta\in\Pi$ is a type symbol and
$\Upsilon\subseteq\terms(\Sigma\cup\Pi)$ is a set of pure type
terms. A pair $\ll\beta,\Upsilon\rl$ in $C$ can be read as
$\sem{\beta}\subseteq\sem{\Upsilon}$.

The functions $\incl$ and $\inclv$ are defined in the following.
Where several alternative definitions of $\inclv$ apply, the first
is used.

\begin{eqnarray*}
\lefteqn{\incl(\tau_1,\tau_2) ~\definedas~ \inclv(\ll\tau_1\rl,\{\ll\tau_2\rl\},\emptyset)}\\
\lefteqn{\inclv(\psi,\Psi,C) ~\definedas~}\\ 
  && \left\{\begin{array}{lr}
    \false & \mbox{if $\Psi=\emptyset$}\\
    \true  & \mbox{if $\psi=\epsilon$} \\
    \multicolumn{2}{l} {\inclv(\tl(\psi),\tls(\Psi),C)} \\
    \multicolumn{2}{r}{
         \mbox{if $\ll\hd(\psi),\Upsilon\rl\in C$ and 
         $\hds(\Psi)\supseteq\Upsilon$}}\\
     \multicolumn{2}{l} {\forall\psi'\in\exp(\psi).
              \inclv(\psi',\Psi,C\cup\{\ll\hd(\psi,\hds(\Psi)\rl\})~~~~}\\
     \multicolumn{2}{r}{\mbox{if $\hd(\psi)\in\Pi$}}\\
     \multicolumn{2}{l}{
       \inclv(\open(\psi),\opens(\sel(\hd(\psi),\exps(\Psi))),C) }\\
      \multicolumn{2}{r}{
             \mbox{if $\hd(\psi)=f(\tau_1,\cdots,\tau_n)$}}     
   \end{array}\right.
\end{eqnarray*}

The second condition $\hds(\Psi)\supseteq\Upsilon$ for the third
alternative is obviously mistaken to be
$\hds(\Psi)\subseteq\Upsilon$ in~\cite{DartZ92}.  The first two
alternatives deal with two trivial cases. The third alternative
uses pairs in $C$ to force termination. As we shall see later,
this is fine for tuple distributive regular types but is
problematic for regular types in general. The fourth alternative
expands $\psi$ into a set of sequences $\psi'$ and compares each
of them with $\Psi$. The fifth alternative applies when
$\psi=f(\tau_1,\cdots,\tau_n)+\psi'$. Sequences in $\Psi$ are
expanded and the expanded sequences of the form
$f(\sigma_1,\cdots,\sigma_n)+\omega'$ are selected. $\psi$ and the
set of the selected sequences are then compared after replacing
$f(\tau_1,\cdots,\tau_n)$ with $\tau_1\cdots\tau_n$ in $\psi$ and
replacing $f(\sigma_1,\cdots,\sigma_n)$ with
$\sigma_1\cdots\sigma_n$ in each
$f(\sigma_1,\cdots,\sigma_n)+\omega'$.

\section{Correctness and Completeness} \label{sec:open}
We now address the correctness and the completeness of Dart-Zobel
algorithm that were left open. We first show that the algorithm is
incorrect for regular types by means of a counterexample.  We then
prove that the algorithm is complete for regular types. Thus, the
algorithm provides an approximate solution to the inclusion
problem of regular types in that it returns true if inclusion
relation holds between its two arguments while the reverse is not
necessarily true.

\subsection{Correctness - a counterexample}
The following example shows that Dart-Zobel algorithm is incorrect
for regular types.
\begin{example} Let $\grammar=\ll\Pi,\Sigma,\rules\rl$ with 
$\Pi=\{\alpha,\beta,\theta,\sigma,\omega\}$,\\ $\Sigma=\{a,b,g(),h(,)\}$ and 
\[ \rules = \left\{\begin{array}{l}
               \alpha\rightarrow g(\omega)\\
               \beta\rightarrow g(\theta)~|~g(\sigma)\\
               \theta\rightarrow a~|~h(\theta,a)\\
               \sigma\rightarrow b~|~h(\sigma,b)\\
               \omega\rightarrow a~|~b~|~h(\omega,a)~|~h(\omega,b)
            \end{array}\right\}   
\]
where, for instance, $\theta\rightarrow a~|~h(\theta,a)$ is an
abbreviation of two rules $\theta\rightarrow a$ and
$\theta\rightarrow h(\theta,a)$. Let
$\Sigma_{h}=\Sigma\setminus\{h\}$. We have
\begin{eqnarray*}
\sem{\theta} &=& \{t\in\terms(\Sigma_{h})~|~\mbox{t is left-skewed and leaves of t are a's}\}\\
\sem{\sigma} &=& \{t\in\terms(\Sigma_{h})~|~\mbox{t is left-skewed and leaves of t are b's}\}\\
\sem{\omega} & = & \{t\in\terms(\Sigma_{h})~|~\mbox{t is left-skewed}\}\\
\sem{\alpha} &=& \{g(t)~|~t\in\sem{\omega}\}\\
\sem{\beta} &=& \{g(t)~|~t\in\sem{\theta}\cup\sem{\sigma}\}
\end{eqnarray*}

Let $t=g(h(h(a,b),a))$. $t\in\sem{\alpha}$ and $t\not\in\sem{\beta}$.
Therefore, $\sem{\alpha}\not\subseteq\sem{\beta}$. The incorrectness
of Dart-Zobel algorithm is illustrated by showing
$\incl(\alpha,\beta)=\true$ as follows. Let
$C_0=\{\ll\alpha,\{\beta\}\rl\}$. We have

\[\begin{array}{lllr}
   \incl(\alpha,\beta) &= & \inclv(\ll\alpha\rl,\{\ll\beta\rl\},\emptyset)
                      & \mbox{by def. of $\incl$} \\
   &=& \inclv(\ll g(\omega)\rl,\{\ll\beta\rl\},C_0) 
     & \mbox{by 4th def. of $\inclv$}\\
   &=& \inclv(\ll\omega\rl,\{\ll\theta\rl,\ll\sigma\rl\},C_0) 
     & \mbox{by 5th def. of $\inclv$}
  \end{array}
\]

Let $C_{1}=C_0\cup\{\ll\omega,\{\theta,\sigma\}\rl\}$. By the fourth
definition of $\inclv$ and the above equation,
\begin{equation} \label{eq:1}
 \incl(\alpha,\beta)=  \left(\begin{array}{lr}
  & \inclv(\ll{a}\rl,\{\ll\theta\rl,\ll\sigma\rl\},C_1)\\ \land &
   \inclv(\ll{b}\rl,\{\ll\theta\rl,\ll\sigma\rl\},C_1)\\ \land &
   \inclv(\ll{h(\omega,a)}\rl,\{\ll\theta\rl,\ll\sigma\rl\},C_1)\\ \land &
   \inclv(\ll{h(\omega,b)}\rl,\{\ll\theta\rl,\ll\sigma\rl\},C_1)
\end{array}\right)
\end{equation}

By applying the fifth and then the second definitions of $\inclv$,
$\inclv(\ll{a}\rl,\{\ll\theta\rl,\ll\sigma\rl\},C_1)=
\inclv(\epsilon,\{\epsilon\},C_1)=\true$. In the same way, we obtain
$\inclv(\ll{a}\rl,\{\ll\theta\rl,\ll\sigma\rl\},C_1)=\true$.

\[\begin{array}{lllr}
\multicolumn{4}{l}{\inclv(\ll{h(\omega,a)}\rl,\{\ll\theta\rl,\ll\sigma\rl\},C_1)}\\ 
  &=&\inclv(\ll{\omega,a}\rl,\{\ll\theta,a\rl,\ll\sigma,b\rl\},C_1) 
  & \mbox{by 5th def. of $\inclv$}\\
  &=& \inclv(\ll{a}\rl,\{\ll a\rl,\ll b\rl\},C_1) 
  & \mbox{by 3rd def. of $\inclv$}\\
  &=& \inclv(\epsilon,\{\epsilon\},C_1) 
  & \mbox{by 5th def. of $\inclv$}\\
  &=& \true
  & \mbox{by 2nd def. of $\inclv$}
  \end{array}
\] 
We can show
${\inclv(\ll{h(\omega,a)}\rl,\{\ll\theta\rl,\ll\sigma\rl\},C_1)}=\true$
in the same way as above. Therefore, by equation~\ref{eq:1},
$\incl(\alpha,\beta)=\true$ and $\incl$ is incorrect for regular types.

{\noindent $\Box$}
\end{example}

The problem with the algorithm stems from the way the set $C$ is used
in the third definition of $\inclv$. As the above example indicates,
the third definition of $\inclv$ severs the dependency between the
terms in a tuple, i.e., subterms of a term.

In~\cite{DartZ92}, Dart and Zobel show by an example that their
algorithm works for some regular types which are not tuple
distributive. We don't know what is the largest subclass of the
class of regular types for which the algorithm is correct.

\subsection{Completeness} 
We now prove that Dart-Zobel algorithm is complete for regular
 types in the sense that $\incl(\tau_1,\tau_2)=\true$ whenever
 $\sem{\tau_1}\subseteq\sem{\tau_2}$.  Let $C$ be a set of pairs
 $\ll\beta,\Upsilon\rl$ with $\beta\in\Pi$ and
 $\Upsilon\subseteq\terms(\Sigma\cup\Pi)$. A pair
 $\ll\beta,\Upsilon\rl$ in $C$ states that the denotation of
 $\beta$ is included in that of $\Upsilon$, i.e.,
 $\sem{\beta}\subseteq\sem{\Upsilon}$ for regular types. Define
\[ \Gamma_{C,\grammar}\definedas \land_{\ll\beta,\Upsilon\rl\in{C}}
	\sem{\beta}\subseteq\sem{\Upsilon}
\]

The completeness of $\incl$ follows from the following theorem which
asserts the completeness of $\inclv$. 

\begin{theorem} \label{th:1}
 Let $\psi$ be a sequence of pure type terms and $\Psi$ a set of
sequences of pure type terms of the same length as $\psi$, $C$ a
set of pairs $\ll\beta,\Upsilon\rl$ with $\beta\in\Pi$ and
$\Upsilon\subseteq\terms(\Sigma\cup\Pi)$. If
$\Gamma_{C,\grammar}\models \sem{\psi}\subseteq\sem{\Psi}$ then
$\inclv(\psi,\Psi,C)=\true$.

\begin{proof}  Assume
$\inclv(\psi,\Psi,C)=\false$. The proof is done by showing
$\Gamma_{C,\grammar}\not\models
\sem{\psi}\subseteq\sem{\Psi}$. This is accomplished by induction
on $\ll{dp}(\psi,\Psi,C),{lg}(\psi)\rl$ where ${lg}(\psi)$ is the
length of $\psi$ and $dp(\psi,\Psi,C)$ is the depth of the
computation tree for $\inclv(\psi,\Psi,C)$. Define $\ll{k},{l}\rl
<\ll{k'},{l'}\rl\definedas (k<k')\lor(k=k')\land(l<l')$.

Basis. ${dp}(\psi,\Psi,C)=0$ and ${lg}(\psi)=0$. $\psi=\epsilon$ and
$\Psi=\emptyset$ since $\inclv(\psi,\Psi,C)=\false$. Let
$\bt=\epsilon$.  $\bt\in\sem{\psi}$ and $\bt\not\in\sem{\Psi}$. So,
$\Gamma_{C,\grammar}\not\models \sem{\psi}\subseteq\sem{\Psi}$.

Induction. ${dp}(\psi,\Psi,C)\neq 0$ or ${lg}(\psi)\neq 0$. By the
definition of $\inclv$,
\begin{itemize}
\item [(a)] $\Psi=\emptyset$; or 
\item [(b)] $\inclv(\tl(\psi),\tls(\Psi),C)=\false$ and there is $\Upsilon\subseteq\terms(\Sigma\cup\Pi)$ such that
            $(\ll\hd(\psi),\Upsilon\rl\in{C})\land(\hds(\Psi)\supseteq\Upsilon)$; or 
\item [(c)] $\hd(\psi)\in\Pi$ and $\exists.\psi'\in\exp(\psi).\inclv(\psi',\Psi,C')=\false$ where $C'=C\cup\{\ll\hd(\psi),\hds(\Psi)\rl\}$; or
\item [(d)] $\hd(\psi)=f(\tau_1,\cdots,\tau_n)$ and
$\inclv(\psi',\Psi',C)=\false$ where $\psi'=\open(\psi)$ and
$\Psi'=\opens(\sel(\hd(\psi),\exps(\Psi)))$.
\end{itemize}

It remains to prove that $\Gamma_{C,\grammar}\not\models
\sem{\psi}\subseteq\sem{\Psi}$ in each of the cases (a)-(d).  The case
(a) is trivial as $\grammar$ is simplified and hence
$\sem{\psi}\neq\emptyset$.

In the case (b), we have
${dp}(\tl(\psi),\tls(\Psi),C)\leq{dp}(\psi,\Psi,C)$ and
${lg}(\tl(\psi))<{lg}(\psi)$. By the induction hypothesis,
$\Gamma_{C,\grammar}\not\models
\sem{\tl(\psi)}\subseteq\sem{\tls(\Psi)}$. Thus,
$\Gamma_{C,\grammar}\models\exists\bt'.(\bt'\in\sem{\tl(\psi)}\land\bt'\not\in\sem{\tls(\Psi)})$. Let
$t\in\sem{\hd(\psi)}$ and $\bt=\ll{t}\rl+\bt'$. Note that $t$ exists
as $\grammar$ is simplified. We have
$\Gamma_{C,\grammar}\models\bt\in\sem{\psi}\land\bt\not\in\sem{\Psi}$.
So, $\Gamma_{C,\grammar}\not\models \sem{\psi}\subseteq\sem{\Psi}$. 

In the case (c), ${dp}(\psi',\Psi,C') < {dp}(\psi,\Psi,C)$. By the
induction hypothesis, $\Gamma_{C'\grammar}\not\models
\sem{\psi'}\subseteq\sem{\Psi}$. Note that
$\Gamma_{C',\grammar}=\Gamma_{C,\grammar}\land(\sem{\hd(\psi)}\subseteq\sem{\hds(\Psi)})$. So,
we have $\Gamma_{C,\grammar}\models
\sem{\psi}\not\subseteq\sem{\Psi}\lor\sem{\hd(\psi)}\not\subseteq\sem{\hds(\Psi)}$. Assume
$\Gamma_{C,\grammar}=\true$.  Either (i)
$\sem{\psi}\not\subseteq\sem{\Psi}$ or (ii)
$\sem{\hd(\psi)}\not\subseteq\sem{\hds(\Psi)}$. In the case (i),
$\exists.\bt'.(\bt'\in\sem{\psi'})\land(\bt'\not\in\sem{\Psi})$.
By proposition 5.26 in~\cite{DartZ92}, we have
$\Gamma_{C,\grammar}\not\models \sem{\psi}\subseteq\sem{\Psi}$. In
the case (ii), $\exists
t.(t\in\sem{\hd(\psi)})\land(t\not\in\sem{\hds(\Psi)})$. Let
$\bt'\in\sem{\tl(\psi)}$ and $\bt=\ll t\rl + \bt'$. Note that
$\bt'$ exists as $\grammar$ is simplified.  We have
$\bt\in\sem{\psi}\land\bt\not\in\sem{\Psi}$. So,
$\Gamma_{C,\grammar}\not\models \sem{\psi}\subseteq\sem{\Psi}$ in
the case (c).

In the case (d), we have $\psi'=\tau_1\cdots\tau_n+\tl(\psi)$ and
${dp}(\psi',\Psi',C)<{dp}(\psi,\Psi,C)$. By the induction hypothesis,
$\Gamma_{C,\grammar}\not\models
\sem{\psi'}\subseteq\sem{\Psi'}$. Thus,
$\Gamma_{C,\grammar}\models\exists\bt_{1}.\exists\bt_{2}.
({lg}(\bt_1)=n)\land((\bt_{1}+\bt_{2})\in\sem{\psi'})\land((\bt_{1}+\bt_{2})\not\in\sem{\Psi'})$,
which implies
$\Gamma_{C,\grammar}\models\exists\bt_{1}.\exists\bt_{2}.  (\ll
f(\bt_{1})\rl+\bt_{2})\in\sem{\psi})\land((\ll
f(\bt_{1})\rl+\bt_{2})\not\in\sem{\Psi})$. So,
$\Gamma_{C,\grammar}\not\models \sem{\psi}\subseteq\sem{\Psi}$.

{\noindent $\Box$}
\end{proof}
\end{theorem}

The completeness of $\incl$ is a corollary of the above theorem. 
\begin{corollary} \label{co:1}
Let $\tau_1$ and $\tau_2$ be pure type terms. If
$\sem{\tau_1}\subseteq\sem{\tau_2}$ then $\incl(\tau_1,\tau_2)=\true$. 

\begin{proof} 
$\incl(\tau_1,\tau_2)=\inclv(\ll\tau_1\rl,\{\ll\tau_2\rl\},\emptyset)$
by the definition of $\incl$. We have
$\Gamma_{\emptyset,\grammar}\models
\sem{\ll\tau_1\rl}\subseteq\sem{\{\ll\tau_2\rl\}}$ since
$\sem{\tau_1}\subseteq\sem{\tau_2}$. The corollary now follows from
the above theorem as $\Gamma_{\emptyset,\grammar}=\true$.

{\noindent $\Box$}
\end{proof}
\end{corollary}

\section{Tuple Distributive Regular Types}\label{sec:tuple}
Most type languages in logic programming use tuple distributive
closures of regular term languages as
types~\cite{Soloman78,Mishra:84,Zobel:87,HeintzeJ90,Jacods:PLDI90,Reddy:NACLP90,YardeniFS:ICLP91,FruhwirthSVY:LICS91,Yardeni:Shapiro:91,HeintzeJ92,GallagherW94,Lu95,CL96,LuJLP98}. The
notion of tuple distributivity is due to
Mishra~\cite{Mishra:84}. The following definition of tuple
distributivity is due to Heintze and
Jaffar~\cite{HeintzeJ90}. Each function symbol of arity  $n$ is
associated with $n$ projection operators
$f^{-1}_{(1)},f^{-1}_{(2)},\cdots,f^{-1}_{(n)}$. Let $S$ be a set
of ground terms in $\terms(\Sigma)$. $f^{-1}_{(i)}$ is defined as
follows.
\[ f^{-1}_{(i)}(S)\definedas \{t_{i}~|~f(t_1,\cdots,t_{i},\cdots,t_n)\in S\}
\] The tuple distributive closure of $S$ is 
\[ S^{\star} \definedas \{c~|~c\in S\land c\in\Sigma_{0}\}
         \cup \{f(t_1,\cdots,t_n)~|~t_{i}\in{(f^{-1}_{(i)}(S))}^{\star}\}
\]
where $\Sigma_{0}$ is the set of constants in $\Sigma$. 

The following proposition results from the fact that ${(.)}^\star$
is a closure operator and preserves set inclusion, i.e.,
$S_1\subseteq S_2$ implies $S_1^\star\subseteq S_2^\star$.

\begin{proposition}\label{lm:1} Let $S_1,S_2\subseteq\terms(\Sigma)$. 
\({(S_1\cup S_2)}^{\star}={(S_1^\star\cup S_2^\star)}^{\star}\).

{\noindent $\Box$}
\end{proposition}

The tuple distributive regular type $\tsem{\tau}$ associated with
a pure type term $\tau$ is the tuple distributive closure of the
regular type $\sem{\tau}$ associated with $\tau$~\cite{Mishra:84}.
\[ \tsem{\tau}\definedas \sem{\tau}^{\star}
\] 

Let $\psi$ be a sequence of pure type terms, $\Psi$ be a set of
sequences of pure type terms of the same length.
\begin{eqnarray*}
   \tsem{\epsilon} & \definedas & \{\epsilon\}\\
   \tsem{\psi} &\definedas & 
       \{\ll t\rl + \bt~|~t\in\tsem{\hd(\psi)}\land\bt\in\tsem{\tl(\psi)}\}\\
  \tsem{\Psi} &\definedas& {(\bigcup_{\psi\in\Psi}\tsem{\hd(\psi)})}^{\star}\times\tsem{\tls(\Psi)}
\end{eqnarray*}
The definition of $\tsem{\Psi}$ makes use of tuple distributivity
and hence severs the inter-dependency between components of a
sequences of terms.

\subsection{Correctness}

We now prove that Dart-Zobel algorithm is correct for tuple
distributive regular types in the sense that if
$\incl(\tau_1,\tau_2)=\true$ then
$\tsem{\tau_1}\subseteq\tsem{\tau_2}$.  Let $C$ be a set of pairs
$\ll\beta,\Upsilon\rl$ with $\beta\in\Pi$ and
$\Upsilon\subseteq\terms(\Sigma\cup\Pi)$. A pair
$\ll\beta,\Upsilon\rl$ in $C$ represents
$\tsem{\beta}\subseteq\tsem{\Upsilon}$ for tuple distributive
regular types. Define
\[ \Phi_{C,\grammar}\definedas \land_{\ll\beta,\Upsilon\rl\in{C}}
	\tsem{\beta}\subseteq\tsem{\Upsilon}
\]

The correctness of $\incl$ follows from the following theorem
which asserts the correctness of $\inclv$ for tuple distributive
regular types.

\begin{theorem} \label{th:2}
 Let $\psi$ be a sequence of pure type terms and $\Psi$ a set of
sequences of pure type terms of the same length as $\psi$, $C$ a
set of pairs $\ll\beta,\Upsilon\rl$ with $\beta\in\Pi$ and
$\Upsilon\subseteq\terms(\Sigma\cup\Pi)$. If
$\inclv(\psi,\Psi,C)=\true$ then $\Phi_{C,\grammar}\models
\tsem{\psi}\subseteq\tsem{\Psi}$.

\begin{proof} 
Assume $\inclv(\psi,\Psi,C)=\true$.  The proof is done by
induction on $\ll{dp}(\psi,\Psi,C),{lg}(\psi)\rl$.

Basis. ${dp}(\psi,\Psi,C)=0$ and ${lg}(\psi)=0$. $\psi=\epsilon$
and $\Psi\neq\emptyset$ by the second definition of $\inclv$. So,
$\Psi=\{\epsilon\}$ and $\Phi_{C,\grammar}\models
\tsem{\psi}\subseteq\tsem{\Psi}$.

Induction. ${dp}(\psi,\Psi,C)\neq 0$ or ${lg}(\psi)\neq 0$ By the
definition of $\inclv$,
\begin{itemize}
\item [(a)] $\inclv(\tl(\psi),\tls(\Psi),C)=\true$ and there is
            $\Upsilon\subseteq\terms(\Sigma\cup\Pi)$ such that
            $(\ll\hd(\psi),\Upsilon\rl\in{C})\land(\hds(\Psi)
            \supseteq\Upsilon)$; or
\item [(b)] $\hd(\psi)\in\Pi$ and $\forall.\psi'\in\exp(\psi).\inclv(\psi',\Psi,C')=\true$ where $C'=C\cup\{\ll\hd(\psi),\hds(\Psi)\rl\}$; or
\item [(c)] $\hd(\psi)=f(\tau_1,\cdots,\tau_n)$ and
$\inclv(\psi',\Psi',C)=\true$ where $\psi'=\open(\psi)$ and
$\Psi'=\opens(\sel(\hd(\psi),\exps(\Psi)))$.
\end{itemize}

It remains to prove that $\Phi_{C,\grammar}\models
\tsem{\psi}\subseteq\tsem{\Psi}$ in each of the cases (a)-(c).

In the case (a), we have
${dp}(\tl(\psi),\tls(\Psi),C)\leq{dp}(\psi,\Psi,C)$ and
${lg}(\tl(\psi))<{lg}(\psi)$. By the induction hypothesis,
$\Phi_{C,\grammar}\models
\tsem{\tl(\psi)}\subseteq\tsem{\tls(\Psi)}$. 
$(\ll\hd(\psi),\Upsilon\rl\in{C})$ and $(\hds(\Psi)\supseteq\Upsilon)$
imply $\Phi_{C,\grammar}\models
\tsem{\hd(\psi)}\subseteq\tsem{\hds(\Psi)}$. Thus,
$\Phi_{C,\grammar}\models \tsem{\psi}\subseteq\tsem{\Psi}$ by the
definitions of $\tsem{\psi}$ and $\tsem{\Psi}$. Note that 
tuple distributivity is used in the definition of $\tsem{\Psi}$.

In the case (b), ${dp}(\psi',\Psi,C') < {dp}(\psi,\Psi,C)$. By the
induction hypothesis, $\Phi_{C',\grammar}\models
\tsem{\psi'}\subseteq\tsem{\Psi}$. Note that
$\Phi_{C',\grammar}=\Phi_{C,\grammar}\land(\tsem{\hd(\psi)}\subseteq\tsem{\hds(\Psi)})$. So,
\[\Phi_{C,\grammar}\models
\tsem{\psi'}\subseteq\tsem{\Psi}\lor\tsem{\hd(\psi)}\not\subseteq\tsem{\hds(\Psi)}\]
$\inclv(\hd(\psi),\hds(\Psi),C)=\true$ since
$\inclv(\psi,\Psi,C)=\true$. We
have
$\Phi_{C,\grammar}\models\tsem{\hd(\psi)}\subseteq\tsem{\hds(\Psi)}$
by the induction hypothesis since $dp(\hd(\psi),\hds(\Psi),C)<dp(\psi,\Psi,C)$. So, $\Phi_{C,\grammar}\models
\tsem{\psi'}\subseteq\tsem{\Psi}$.  $\Phi_{C,\grammar}\models
(\tsem{\{\psi'~|~\psi'\in\exp(\psi)\}}\subseteq\tsem{\Psi})$ since
${(.)}^\star$ is a closure operator and hence
$\Phi_{C,\grammar}\models \tsem{\psi}\subseteq\tsem{\Psi}$

In the case (c), we have $\psi'=\tau_1\cdots\tau_n+\tl(\psi)$ and
${dp}(\psi',\Psi',C)<{dp}(\psi,\Psi,C)$. By the induction
hypothesis, $\Phi_{C,\grammar}\models
\tsem{\psi'}\subseteq\tsem{\Psi'}$. By proposition 5.29
in~\cite{DartZ92}, $\Phi_{C,\grammar}\models
\tsem{\psi}\subseteq\tsem{\Psi}$. This completes the proof of the
theorem.

{\noindent $\Box$}
\end{proof}
\end{theorem}

The correctness of $\incl$ is a corollary of the above theorem.
\begin{corollary} 
Let $\tau_1$ and $\tau_2$ be pure type terms. If
$\incl(\tau_1,\tau_2)=\true$ then
$\tsem{\tau_1}\subseteq\tsem{\tau_2}$.

\begin{proof} Let $\incl(\tau_1,\tau_2)=\true$. 
$\inclv(\ll\tau_1\rl,\{\ll\tau_2\rl\},\emptyset)=\true$ by the
definition of $\incl$.  Thus, $\Phi_{\emptyset,\grammar}\models
\tsem{\ll\tau_1\rl}\subseteq\tsem{\{\ll\tau_2\rl\}}$ according to
the above theorem. So, $\tsem{\tau_1}\subseteq\tsem{\tau_2}$ as
$\Phi_{\emptyset,\grammar}=\true$.

{\noindent $\Box$}
\end{proof}
\end{corollary}

\subsection{Completeness} 
This section presents the completeness of Dart-Zobel algorithm for
tuple distributive regular types. The following theorem is the
counterpart of theorem~\ref{th:1}.
 
\begin{theorem} \label{th:6}
Let $\psi$ be a sequence of pure type terms and $\Psi$ a set of
sequences of pure type terms of the same length as $\psi$, $C$ a
set of pairs $\ll\beta,\Upsilon\rl$ with $\beta\in\Pi$ and
$\Upsilon\subseteq\terms(\Sigma\cup\Pi)$. If
$\Phi_{C,\grammar}\models \tsem{\psi}\subseteq\tsem{\Psi}$ then
$\inclv(\psi,\Psi,C)=\true$.

\begin{proof} 
 The proof can be obtained from that for theorem~\ref{th:1} by
 simply replacing $\Gamma_{\cdot,\cdot}$ with $\Phi_{\cdot,\cdot}$
 and $\sem{{\cdot}}$ with $\tsem{{\cdot}}$.

{\noindent $\Box$}
\end{proof}
\end{theorem}

The following completeness result of Dart-Zobel algorithm for
tuple distributive regular types follows from the above theorem.

\begin{corollary} 
Let $\tau_1$ and $\tau_2$ be pure type terms. If
$\tsem{\tau_1}\subseteq\tsem{\tau_2}$ then $\incl(\tau_1,\tau_2)=\true$. 

\begin{proof} 
 The proof can be obtained from that for corollary~\ref{co:1} by
 simply replacing $\Gamma_{(\cdot,\cdot)}$ with
 $\Phi_{(\cdot,\cdot)}$, $\sem{{\cdot}}$ with $\tsem{{\cdot}}$ and 
 theorem~\ref{th:1} with theorem~\ref{th:6}.

{\noindent $\Box$}
\end{proof}
\end{corollary}

\subsection{A Simplified Algorithm}
Now that Dart-Zobel algorithm is complete and correct for tuple
distributive regular types but not correct for general regular
types. It is desirable to specialise Dart-Zobel algorithm for
tuple distributive regular types which was originally proposed for
general regular types. The following is a simplified version of
the algorithm for tuple distributive regular types.

\begin{eqnarray*}
\lefteqn{\incl'(\tau_1,\tau_2) ~\definedas~ \incl'(\tau_1,\{\tau_2\},\emptyset)}\\
\lefteqn{\incl'(\tau,\Upsilon,C) ~\definedas~}\\ 
  && \left\{\begin{array}{lr}
    \false & \mbox{if $\Upsilon=\emptyset$}\\
    \multicolumn{2}{l}{\true}\\
    \multicolumn{2}{r}{\mbox{if $(\ll\tau,\Upsilon'\rl\in C)\land(\Upsilon\supseteq\Upsilon')$}}\\
    \forall\tau'\in\exp'(\tau).
              \incl'(\tau',\Upsilon,C\cup\{\ll\tau,\Upsilon\rl\})
       &\mbox{if $\tau\in\Pi$}\\
     \multicolumn{2}{l}{
       \inclv'(\tau_1\cdots\tau_n,
              \{\sigma_1\cdots\sigma_n~|~f(\sigma_1,\cdots,\sigma_n)\in\exps'(\Upsilon)\},C)~~~~ }\\
      \multicolumn{2}{r}{
             \mbox{if $\tau=f(\tau_1,\cdots,\tau_n)$}}     
   \end{array}\right.\\
\lefteqn{\inclv'(\epsilon,\{\epsilon\},C) ~\definedas~ \true}\\
\lefteqn{\inclv'(\psi,\Psi,C) ~\definedas~}\\
  &&~~~\incl'(\hd(\psi),\hds(\Psi),C)\land  
   \inclv'(\tl(\psi),\tls(\Psi),C)\\
\lefteqn{ \exp'(\tau) \definedas 
    \left\{\begin{array}{ll} \{\tau\} & \mbox{if
     $\tau\not\in\Pi$}\\
     \{\sigma~|~\tau\rightarrow\sigma)\in\rules\} &
     \mbox{if $\tau\in\Pi$}\\ \end{array}\right.}\\
\lefteqn{ \exps'(\Upsilon)\definedas\bigcup_{\tau\in\Upsilon}\exp'(\tau)}
\end{eqnarray*}

While Dart-Zobel algorithm mainly deals with sequences of pure
type terms, the simplified algorithm primarily deals with pure
type terms by breaking a sequence of pure type terms into its
component pure type terms. This is allowed because tuple
distributive regular types abstract away inter-dependency between
component terms in a sequence of ground terms. We forgo presenting
the correctness and the completeness of the simplified algorithm
because they can be proved by emulating proofs for
theorems~\ref{th:1} and~\ref{th:2}.

\section{Conclusion}\label{sec:conc} 
We have provided answers to open questions about the correctness
and the completeness of Dart-Zobel algorithm for testing inclusion
of one regular type in another. The algorithm is complete but
incorrect for general regular types. It is both complete and
correct for tuple distributive regular types. It is our hope that
the results presented in this paper will help identify the
applicability of Dart-Zobel algorithm. We have also provided a
simplified version of Dart-Zobel algorithm for tuple distributive
regular types.


\end{document}